\documentclass[twoside,fleqn]{article}
\usepackage{espcrc2}
\usepackage{epsfig}
\usepackage{psfig}
\newcommand{\beqn} {\begin{equation}}
\newcommand{\eqn} {\end{equation}}

\def\lsim{\raise0.3ex\hbox{$<$\kern-0.75em\raise-1.1ex\hbox{$\sim$}}}
\def\gsim{\raise0.3ex\hbox{$>$\kern-0.75em\raise-1.1ex\hbox{$\sim$}}}

\newcommand{\ep}{\epsilon}
\newcommand{\ga}{\gamma}
\newcommand{\al}{\alpha}
\newcommand{\be}{\beta}

\newcommand{\AmS}{{\protect\the\textfont2
  A\kern-.1667em\lower.5ex\hbox{M}\kern-.125emS}}

\title{
\vskip -100pt
\mbox{} \hfill BI-TP 98/25\\
\mbox{} \hfill September 1998\\
\vskip 45pt
Constituent Quarks, Diquarks and the $N-\Delta$ Mass Splitting
\thanks{Talk given at the XVI International Symposium on Lattice Field
Theory, Boulder, July 13-18,1998}
}
\author{F. Karsch with M. He{\ss}, E. Laermann and I. Wetzorke
\\
\vskip 6pt
Fakult\"at f\"ur Physik, Universit\"at Bielefeld,
D-33615 Bielefeld, Germany}
\begin{document}
\begin{abstract}
We analyze hadron as well as quark and diquark correlation functions in Landau 
gauge in order to extract information on the spin dependence of the quark-quark 
interaction. We find evidence that the $N-\Delta$ mass splitting can be 
attributed to the spin dependence of the interaction between quarks in a colour 
anti-triplet state with spin 0 and 1, respectively. The lightest excitations are
observed in the $S=0$ channel. However, no evidence for a deeply bound 
diquark state is found.

\end{abstract}

\maketitle
\vskip 20pt
%
\noindent
\section{INTRODUCTION}
The spin dependent interaction among quarks found in perturbation
theory from one gluon exchange diagrams (OGE) \cite{deR75} as well as induced
by instantons leads at least to a qualitative understanding of the 
fine structure of the hadron spectrum \cite{Glo96}. 
In particular, the attractive interaction in the $S=0$ channel has led to the
idea that diquarks may be well defined, localized objects (bound states?)
within a nucleus. This also led to the speculation that diquarks may play
an important role for the phase structure of QCD at high density. The 
possibility of a diquark gas \cite{Don88} has been discussed 
and recently new ideas about the possible existence of a 
diquark condensate ({\it colour superconductor}) \cite{Ying,Rap97,Alf98} 
have been put forward.

Whether these newly proposed phases of dense matter are realized in nature
depends on details 
of the spin dependent part of the q-q interaction. Also
in the case of the hadron spectrum it is of interest to analyze the relative
importance of the different mechanisms that can give rise to the observed
spin splitting of hadronic states. Already in the case of the $N-\Delta$ 
mass splitting the contributions from $S=0$ and 1 channels differ in the case of
OGE and instanton induced interactions. Both lead to attractive terms
for spin 0 diquarks in a colour anti-triplet ($\bar{3}_c$) state they predict,
however, opposite signs for the interaction in the $S=1$ channel (see Table 1).   

\begin{table}
\begin{center}
\begin{tabular}{|l|c|r|r|}\hline
$(F,S,C)$&state&Inst.&OGE  \\
\hline
\hline
$(\bar{3},0,\bar{3})$&
$\ep_{abc} (C\ga_5)_{\al\be}u_{a,\al}^{\dagger}d_{b,\be}^{\dagger}$
&-2&-2 \\
$(6,1,\bar{3})$&$\ep_{abc}u_{a,\alpha}^{\dagger}u_{b,\alpha}^{\dagger}$
&-1/3&2/3 \\
$(\bar{3},1,6)$&$u_{c,\alpha}^{\dagger}d_{c,\alpha}^{\dagger}$
&2/3&-1/3 \\
$(6,0,6)$&$(C\ga_5)_{\al\be}u_{c,\al}^{\dagger}u_{c,\be}^{\dagger}$
&1&1 \\
\hline
\end{tabular}
\end{center}
\caption{Diquark states with spin $S$ and different flavour ($F$) and colour
($C$) representations. The last two 
columns give the relative strength of interaction terms corresponding to
a flavour-spin (instanton) and colour-spin (OGE) 
coupling, $V_S \sim (\lambda_1^a \lambda_2^a)
(s_1 s_2)$, where $\lambda_i^a$ denote $SU(3)$ generators.}
\vspace*{-0.5truecm}
\label{tab:states}
\end{table}

While calculations within the instanton liquid model gave indications for
a quite light (deeply bound) diquark state \cite{Sch94} an analysis of the
quark distribution inside a nucleus did not give any hints for well 
localized diquarks \cite{Lei93}. In the following we will present some
results from our analysis of the q-q interaction performed in Landau 
gauge within the quenched approximation of QCD \cite{paper}. 

The calculations we are reporting here are based on an analysis of 73
gauge field configurations, generated on $16^3\times 32$ lattices at 
$\beta=4.1$ with a tree-level Symanzik improved action.
After fixing the Landau gauge fermionic correlation 
functions have been calculated with four different source vectors using a 
tree-level improved clover action. Propagators have been analyzed for 8
values of the hopping parameter, $\kappa \in [0.14,0.148]$, which corresponds
to the hadron mass interval $0.5 < m_\pi / m_\rho < 0.9$. From a calculation
of the string tension we find the lattice cut-off, $a^{-1} \simeq 1.1$~GeV; 
our largest $\kappa$-value corresponds to a pion mass 
$m_\pi = (0.316\pm 0.003 ) a^{-1} \simeq 350$~MeV. 

\section{$N-\Delta$ MASS SPLITTING}

On our data set we have calculated correlation functions for the 
nucleon and delta. We find that the $N-\Delta$ mass difference 
increases with decreasing quark massi. As shown in Fig.~1 results are
also quantitatively
consistent with earlier findings from 
calculations with Wilson fermions \cite{qcdpax}.
\begin{figure}[t]
\epsfig{file=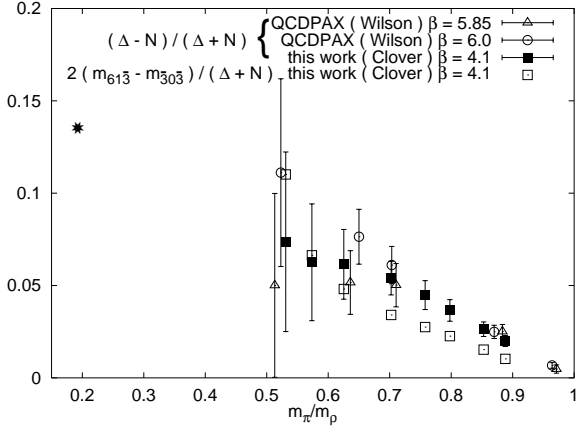,width=80mm}
\vskip -0.7truecm
\caption{The $N-\Delta$ mass splitting versus the pion-rho mass ratio. 
The star denotes the experimental value.
Data for Wilson fermions are taken from [10]. Also shown is the
mass splitting between $S=0$ and 1 diquarks (open squares). Errors are
similar to those shown for the $N-\Delta$ mass splitting.}
\label{fig:split}
\vskip -0.5truecm
\end{figure}

Simple constituent quark models relate the $N-\Delta$ mass splitting to the
spin dependence of the interaction among quarks in a $\bar{3}_c$ state,
{\it i.e.} $m_\Delta - m_N = 2 (V_{S=1} -V_{S=0})$.
In order to test in how far such a relation can hold we have calculated 
diquark correlation functions in Landau gauge,
\begin{equation}
G_{F,S,C} (t) = \langle D(0) D^{\dagger} (t) \rangle~~.
\end{equation}
Here $D(t)$ denotes one of the states given in the second column of
Table~1. 
At large distances the correlation functions are expected to decay 
exponentially with a mass which is characteristic for the given quantum number
channel. 
In Fig.~2 we show the ratio of correlation functions for 
$\bar{3}_c$ diquarks with $S=0$ and $1$,  
\begin{equation}
G_{\bar{3}0\bar{3}}(t)~ /~ G_{61\bar{3}}(t) \sim {\rm exp} {\bigl(
[m_{61\bar{3}} - m_{\bar{3}0\bar{3}}]\ t \bigr)}~~. 
\end{equation}
\begin{figure}[t]
\epsfig{file=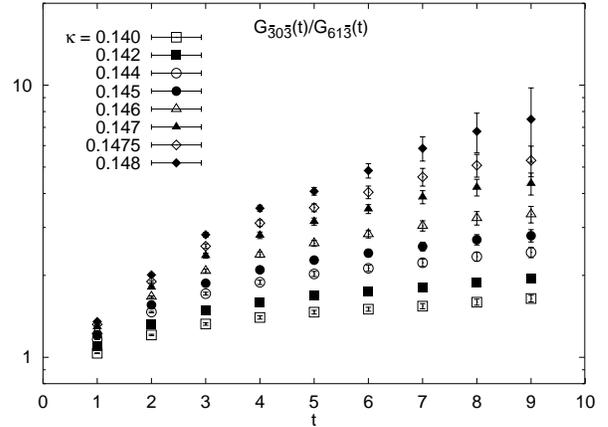,width=80mm}
\vskip -0.7truecm
\caption{Ratio of correlation functions of colour anti-triplet
states with spin $S=0$ and $S=1$ versus Euclidean time $t$.}  
\label{fig:tt}
\vskip -0.4truecm
\end{figure}
The ratio of correlation function clearly rises exponentially at large 
distances; the slope increases
with decreasing quark mass. The mass differences extracted from 
fits to this correlation functions give, again within simple potential models,
the difference between the spin-dependent q-q interactions,
$m_{61\bar{3}} - m_{\bar{3}0\bar{3}} \equiv (V_{S=1} -V_{S=0})$. Twice
this difference is shown in Fig.~1 (open squares). As can be seen it is 
quite compatible, although systematically below the results for 
the $N-\Delta$ mass splitting.

\section{OGE VERSUS INSTANTONS} 

In order to differentiate between predictions of OGE and
instanton induced interaction models for different quantum number channels
one may compare the $S=1$ correlators in $\bar{3}_c$ and $6_c$ channels.
The ratio $G_{\bar{3}16}(t) / G_{61\bar{3}}(t)$ 
is expected to rise (fall) with increasing $t$ in the 
case of OGE (instanton) induced q-q interactions (Table~1). From the results
shown in Fig.~3 it is obvious that at
least for heavy quarks the interaction
\begin{figure}[t]
\epsfig{file=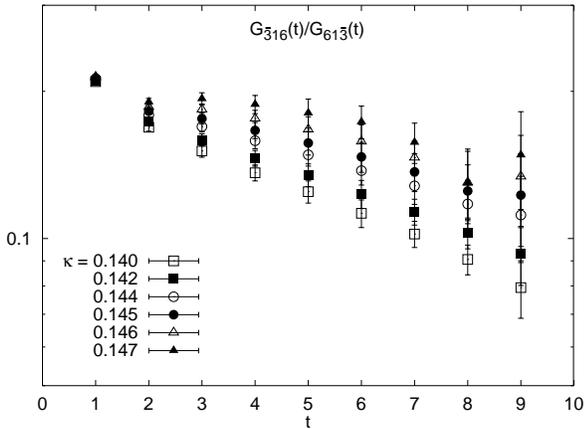,width=80mm}
\vskip -0.7truecm
\caption{Ratio of correlation functions of $S=1$, colour $\bar{3}$ and $6$ 
states versus Euclidean time $t$. 
}
\label{fig:sextet}
\vskip -0.7truecm
\end{figure}
in the $\bar{3}_c$ channel leads to lighter excitations, {\it i.e.} is
more attractive. This is in favour of the instanton induced interaction models.
We note, however, that with decreasing quark mass there is a tendency for
the ratios to flatten. A $t$-independent ratio in the chiral limit thus
cannot be ruled out. This could suggest that no bound states exist in these
quantum number channels. 

\section{QUARK AND DIQUARK MASSES}

The above discussed analysis suggests that the diquark correlator in the
$\bar{3}_c$, spin zero channel leads to the lightest excitations. 
The correlation function $G_{\bar{3}0\bar{3}} (t)$ does indeed show
a rather clean exponential decay, which leads to a plateau for local masses
at distances $t\sim 4$ for all quark masses considered by us.
This suggests that the diquark is a well localized state. Using the string tension
to set the scale ($\sqrt{\sigma}=420$MeV) we find from an extrapolation to the 
chiral limit $m_{\bar{3}0\bar{3}} = 694(22)$MeV.
 
A similarly stable behaviour is found for local masses extracted from
the quark propagator. Unlike for all diquark masses the plateau is, 
however, reached from below which reflects the non-existence of a positive
transfer matrix in Landau gauge. In the chiral limit we find for the constituent 
mass, $m_q = 342(13)$MeV. The mass of the lightest diquark obviously is 
consistent with twice the constituent quark mass. There thus is no evidence 
for a deeply bound diquark state as it has been found in instanton
model calculations. A comparison of quark, diquark and hadron masses is given in 
Fig.~4. As can be seen 1/3 of the nucleon mass lies below the quark mass which 
can be addressed to the contribution of an attractive q-q interaction. 

\begin{figure}[t]
\epsfig{file=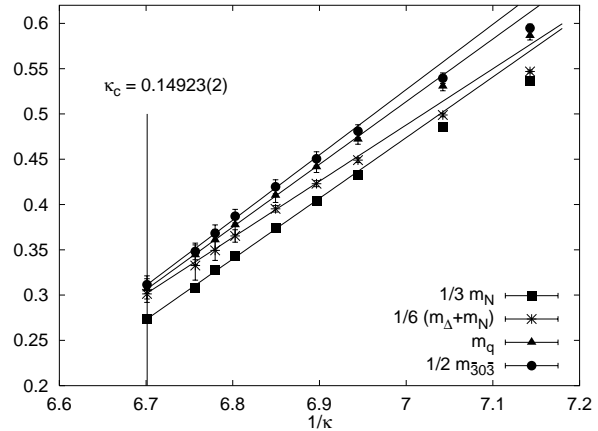,width=80mm}
\vskip -0.7truecm
\caption{Masses in units of the number of valence quarks for 
different values of $\kappa$.}
\label{fig:bound}
\vskip -0.7truecm
\end{figure}

The analysis of diquark correlation functions in different quantum number
channels shows that correlations are stronger in colour anti-triplet channels
than in colour sextet channels. 
Nonetheless there is so far no evidence for a deeply bound 
diquark state. One should, however, stress that
this first exploratory lattice study has been performed in the quenched
approximation on quite coarse lattices and still with fairly large quark masses. 
The influence of these approximations should be analyzed in more detail 
in the future.

\end{document}